\documentclass[preprint]{aastex}
\slugcomment{Submitted to ApJL}
\shortauthors{Zheng et al.}
\begin{document}

%\title{{\bf Jet-coronal hole collision and a closely-related coronal mass ejection}}
\title{An extreme ultraviolet wave generating upward secondary waves in a streamer-like solar structure}
\author{Ruisheng Zheng, Yao Chen, Shiwei Feng, Bing Wang, and Hongqiang Song}
\affil{Shandong Provincial Key Laboratory of Optical Astronomy and Solar-Terrestrial Environment, and Institute of Space Sciences, Shandong University, 264209 Weihai, China; ruishengzheng@sdu.edu.cn\\}

\begin{abstract}
Extreme ultraviolet (EUV) waves, spectacular horizontally propagating disturbances in the low solar corona, always trigger horizontal secondary waves (SWs) when they encounter ambient coronal structure. We present a first example of upward SWs in a streamer-like structure after the passing of an EUV wave. The event occurred on 2017 June 1. The EUV wave happened during a typical solar eruption including a filament eruption, a CME, a C6.6 flare. The EUV wave was associated with quasi-periodic fast propagating (QFP) wave trains and a type II radio burst that represented the existence of a coronal shock. The EUV wave had a fast initial velocity of $\sim$1000 km s$^{-1}$, comparable to high speeds of the shock and the QFP wave trains. Intriguingly, upward SWs rose slowly ($\sim$80 km s$^{-1}$) in the streamer-like structure after the sweeping of the EUV wave. The upward SWs seemed to originate from limb brightenings that were caused by the EUV wave. All the results show the EUV wave is a fast-mode magnetohydrodynamic shock wave, likely triggered by the flare impulses. We suggest that part of the EUV wave was probably trapped in the closed magnetic fields of streamer-like structure, and upward SWs possibly resulted from the release of trapped waves in the form of slow-mode. It is believed that an interplay of the strong compression of the coronal shock and the configuration of the streamer-like structure is crucial for the formation of upward SWs.

\end{abstract}

\keywords{Sun: activity --- Sun: corona --- Sun: coronal mass ejections (CMEs)}

\section{Introduction}
%Coronal heating problem of how the outer atmosphere of the Sun is heated to million-degree temperatures, remains one of the most important unsolved problems in Solar Physics and Space Science~\cite{klim06, parn12, klim15}. Currently there are two classes of theories: the wave model and the nanoflare model~\cite{parker88}. The solar atmosphere is much more dynamics, and many transient activities can be related to the corona heating, such as EUV cyclones and the interaction between active regions (ARs) and the quiet Sun~\cite{zhang11, zhang15}. It is believed that a variety of waves and wave-like phenomena in the corona play important roles in the process of corona heating~\cite{ionson78, heyv83, taro09, vanball11}.

Extreme ultraviolet (EUV) waves are well known as horizontal propagating disturbances in the solar corona, and are always associated with energetic eruptions such as coronal mass ejections (CMEs) and flares (Wills-Davey \& Attrill 2009, Gallagher \& Long 2011; Nitta et al. 2013; Liu \& Ofman 2014; Warmuth 2015). The fast EUV waves are frequently accompanied by type II radio bursts (Muhr et al. 2010; Ma et al. 2011; Kozarev et al. 2011; Nitta et al. 2013, 2014), which reveals the formation of coronal magnetohydrodynamic (MHD) shocks. Since its discovery, the EUV wave has been strongly debated on its physical nature (Thompson et al. 1998; Chen et al. 2002; Zhukov \& Auch\`{e}re 2004; Ballai et al. 2005; Attrill et al. 2007; Wang et al. 2009; Zheng et al. 2012). Benefiting with high quality observations from Atmospheric Imaging Assembly (AIA; Lemen et al. 2012) on Solar Dynamics Observatory (SDO; Pesnell et al. 2012), the EUV wave has been best described as a bimodal composition of an outer fast-mode MHD wave and an inner non-wave CME component (Liu et al. 2010; Chen \& Wu 2011; Downs et al. 2012). It is generally believed that the onset of EUV waves strongly depends on the lateral expansion of CME flank and is weakly related to flares (Patsourakos \& Vourlidas 2012; Liu \& Ofman 2014). As well as EUV waves, coronal shocks are almost always excited by CMEs, and only a few cases of shocks are initiated by flares (Vr{\v s}nak et al. 2006; Vr{\v s}nak \& Cliver 2008; Magdaleni{\'c} et al. 2012; Luli{\'c} et al. 2013).

When EUV waves encounter quasi-separatrix layers (QSLs; Delann{\'e}e \& Aulanier 1999) or magnetic separatrixs (e.g. coronal holes (CHs), coronal cavities, and streamers), there usually appear reflections, refractions, and/or transmissions (Li et al. 2012; Olmedo et al. 2012; Shen et al. 2013) in the form of horizontal secondary waves (SWs). On the other hand, EUV waves may sometimes trigger stationary brightenings at structural boundaries of QSLs (Ofman \& Thompson 2002; Chandra et al. 2016). Stationary brightenings resulted from the compression or/and persistent heating in wave or non-wave mechanisms (Ofman \& Thompson 2002; Attrill et al. 2007; Delann{\'e}e et al. 2007). Recently, it is proposed that stationary brightenings are as a result of the wave conversion from fast-mode to slow-mode, after the EUV wave is partially trapped inside the closed magnetic loops (Chen et al. 2016; Zong \& Dai 2017).

Almost all of the SWs in previous studies moved horizontally over the solar surface (Li et al. 2012; Olmedo et al. 2012; Shen et al. 2013). Here we present a first example of upward SWs in a streamer-like structure after the sweeping of an EUV wave. We organize the Letter as follows. In Section 2 we describe the observations used in this work; Then the main results are reported in Section 3; Finally, conclusions and discussion are given in Section 4.

\section{Observations}
The EUV wave occurred in NOAA AR 12661 ($\sim$N07E89) on 2017 June 1, and was intimately associated with a solar eruption that consisted of a filament eruption, a CME of $\sim$520 km s$^{-1}$, a C6.6 flare, and a type II radio burst. We principally employ the observations from the AIA on SDO to get essential details of the eruption. The AIA instrument's ten EUV and UV wavelengths involve a wide range of temperatures. The AIA images cover the full disk (4096~$\times$4096 pixels) of the Sun and up to 0.5 $R_\odot$ above the limb, with a pixel resolution of 0.6$"$ and a cadence of 12 s. The filament eruption is also confirmed by H$\alpha$ filtergrams from Global Oscillation Network Group (GONG) of National Solar Observatory. The evolution of the CME and streamer(-like) structures in the high corona is captured by Large Angle and Spectrometric Coronagraph (LASCO; Brueckner et al. 1995) onboard SOHO spacecraft. In addition, the associated type II radio burst is recorded by the metric spectrometer from Yunnan Astronomical Observatories (YNAO; Gao et al. 2014) and Learmonth (LEAR; Kennewell \& Steward 2003), in the frequency range of 70-700 MHz and 25-180 MHz, respectively.

%The kinematics of the filament eruption, the EUV wave, and the secondary waves, as shown in Fig.~\ref{f1}, Fig.~\ref{f2} and Fig.~\ref{f4}, are obtained by the time-slice approach. The associated speeds are calculated by linear fits, assuming that the measurement uncertainty of the selected points is 4 pixel ($\sim1.74$ Mm).

\section{Results}
\subsection{Solar Eruption}
The solar eruption is captured in AIA EUV images and the GONG H$\alpha$ filtergram (Figure 1 and the online animation). Note that there was a streamer-like structure involving a coronal cavity at the east limb to the north of AR 12661 (white arrows in Fig.1(a)). A CH around the North Pole extended to the streamer-like structure (black arrows in Fig.1(a)). In the beginning, the erupting filament exhibited the distinct helical structure off the limb in the H$\alpha$ and in AIA EUV passbands (arrows in Fig.1(b)-(d)), indicating the presence of a multithermal magnetic flux rope. During the ascending, the filament untwisted and split into two major parts (arrows in Fig.1(e)). After the eruption, the post-eruption loops (PELs) appeared and shrank downwards in the higher corona (the arrow in Fig.1(f)), and there formed a current sheet (CS) and a cusp structure above the eruption center (Fig.1(g)). In time-space plots (Fig.1(h)-(i)) along the dashed line S1 in Fig.1(f), the speed of the erupting filament was $\sim$450 km s$^{-1}$, and the shrinking velocity of PELs was $\sim$100 km s$^{-1}$.

In composite images (Fig.2(a)-(b)) of Lasco C2 and AIA 193~{\AA}, AR 12661 situated near the north leg of a bright steamer belt, and the streamer-like structure on the north connected with the overlying streamer chains (Fig.2(a)). Due to the perturbation of the CME, the streamer chains became very bright (Fig.2(b)). In the light curve of cusp region in 94~{\AA} (the box in Fig.1(g)), the flare impulsive phase started at $\sim$01:39 UT, and peaked at $\sim$01:51 UT (dotted vertical lines in Fig.2(c)). The radio spectrum of YNAO and LEAR shows well the related type II radio burst that indicates the formation of a coronal shock, and two strong stripes are correspond to the fundamental (F) and harmonic (H) branches (Fig.2(d)). The burst was initiated at $\sim$01:43:30 UT, and F and H branches almost disappeared at $\sim$01:45 and $\sim$01:48 UT. Assuming an initial shock speed of 1500 km s$^{-1}$, we calculated the fitting curves of F and H branches by using the one-fold (red solid), three-fold (red dashed), and five-fold (red dotted) Newkirk coronal density models, respectively (Newkirk 1961). It is obvious that the curves of one-fold Newkirk model fit best with the radio burst. On the other hand, the deduced start height (1.18 $R_\odot$) for the one-fold Newkirk model is also consistent with that of the CME front at the same time (the blue triangle in Fig.3(b)). Furthermore, $\sim$2 minutes after the disappearance of the H branch, another type II radio burst appeared at $\sim$90 MHz, and slowly drifted to $\sim$60 MHz till $\sim$01:55 UT (the yellow box in Fig.2(d)). It is likely a result of an interaction between the coronal shock and the streamer(-like) structure (Feng et al. 2012, 2013; Kong et al. 2013).

\subsection{EUV Wave}
Following the eruption, some wave trains, emanating from the core region of the flare, are clearly seen in running-ratio-difference AIA images (Figure 3 and the online animation). The wave trains were most prominent in 171~{\AA}, indicating a possible responsible plasma temperature of 0.8 Mk (Liu et al. 2012). The wave trains propagated outwards between the AR and the streamer-like structure along coronal loops rooting at the core region of the flare (dashed lines in Fig.3(a)-(b) and the black arrow in Fig.3(e)). In time-space diagrams (Fig.3(g)-(h)), there are four successive wave trains (red, white, green, and black dotted lines) in $\sim$10 minutes ($\sim$01:42-01:52 UT) in 171~{\AA}, suggesting an average period of 2.5 minutes, while only the first and the third wave trains were seen in 193~{\AA}. The wave trains had a speed range of 1200-1600 km s$^{-1}$. The first wave train abruptly vanished when it caught the CME front (blue arrows in Fig.3(c)-(d)) that was slowly expanding at a velocity of $\sim$180 km s$^{-1}$ (blue dotted lines in Fig.3(g)-(h)), and other wave trains travelled with shorter distances. The wave trains are weaker than that in previous studies, but their observational characteristics (the speed, the period, and the disappearance) are consistent with that of quasi-periodic fast propagating (QFP) waves behind CME (Liu et al. 2011, 2012; Shen \& Liu 2012). On the other hand, the top of the coronal cavity became bright (the white arrow in Fig.3(e)), and the coronal loops over the coronal cavity horizontally shifted (red arrows in Fig.3(c) and (e)), which is likely due to the compression of the newly-formed propagating EUV wave (green arrows in Fig.3(d) and (f)).

The EUV wave was distinct in base-ratio-difference AIA images (Figure 4 and the online animation). The moving front of the EUV wave was strong in the northern hemisphere, but was very faint in the southern hemisphere (Fig.4(a)-(f)). Hence, we mainly focus on the northward propagation of the EUV wave. The wave front simultaneously moved on the disk and in the low corona (arrows in Fig.4(a)-(c)). After the passing of the wave, coronal loops above the coronal cavity became bright, especially in 171~{\AA} (red arrows in Fig.4(d)-(f)). The kinematics evolution of the wave front is clearly shown in time-space plots (Fig.4(g)-(i)) along the selected arcs (S3-S4 in Fig.4(a)). S3 starts from the altitude of 0.15 $R_\odot$ off the limb in the due east, and S4 extends northwards from the north boundary of the eruption region to the elongated CH on the disk. The EUV wave front was dark in 171~{\AA} (the white arrow in Fig.4(g)) and bright in 193 and 211~{\AA}, suggestive of heating from 0.8 to 2.0 Mk (Liu et al. 2010, 2012). Nitta et al. (2013) showed that this is a common feature of EUV waves. The following brightenings in 171~{\AA} (the blue arrow in Fig.4(g)) are consistent with the lightened loops (red arrows in Fig.4(d)-(f)). Attractively, the EUV wave had a faster initial velocity of $\sim$1000 km s$^{-1}$ from $\sim$01:43 UT (red dotted lines in Fig.4(h)-(i)), and abruptly decreased to $\sim$380 km s$^{-1}$ from $\sim$01:45 UT (green dashed lines in Fig.4(g)-(i)). The position of rapid deceleration along S3 locates at the south boundary of the lightened loops above the coronal cavity (green triangles in Fig.4(d)-(f)). Following the deceleration, the EUV wave slowly stopped (horizontal lines in Fig.4(h)-(i)). The terminal points along S3 and S4 are correspond to the north end of the streamer-like structure and the south boundary of the elongated CH, respectively (green and blue squares in Fig.4 and Fig.5).
%On the other hand, there was another slower component of $\sim$150 km s$^{-1}$ behind the coronal wave in the lower corona (blue dashed line in Fig.4(h)), which is likely the loop shift due to the wave compression.

\subsection{Secondary Waves}
Intriguingly, upward SWs rose after the passing of the EUV wave, clearly shown in base-ratio-difference images in 193~{\AA} (Figure 5 and the online animated version of Figure 4). The SWs (red arrows in Fig.5(a)-(c)) slowly ascended in sequence from south to north in the streamer-like structure (blue arrows in Fig.5(b)). Furthermore, another upward SW clearly propagated along the loops close to the north boundary of the streamer-like structure (the red arrow in Fig.5(d)). The coronal loops at the north boundary of the streamer-like structure became bright (the black arrow in Fig.5(d)), consistent with the brightenings at the stop position of the EUV wave along S3 (the green dotted line in Fig.4(h)). The kinematics of upward SWs are analysed in time-space plots (Fig.5(e)-(g) and (i)) along some slices in the streamer-like structure (S5-S7, and S9; Fig.5(a) and (d)). The SWs propagated at speeds of $\sim$90-120 km s$^{-1}$, indicated by red dotted lines in Fig.5(e)-(g) and (i). The onsets of SWs are closely associated with the arrival of the EUV wave (green arrows in Fig.5(e)-(g) and (i)). The SWs are traced back to limb brightenings (black arrows in Fig.5(e)-(g)). The limb brightenings were triggered by the EUV wave and quickly disappeared as the release of upward SWs. In addition, there was no signal of upward SWs along the slice (S8; the pink dashed line in Fig.5(c) starting from the interior of the CH (the pink plus in Fig.1(a)) beyond the north end of the streamer-like structure (Fig.5(h)).
%The coronal wave had a falling slope that indicated an arrival delay at lower heights (Liu et al. 2012), likely because that the eruption occurred at the some height off the limb. Note that stationary brightenings appeared on the disk as soon as the arrival of EUV wave, and was replaced by dimmings following the SWs (green arrows in Fig.5(e)-(h)).

\section{Conclusions and Discussion}
We show an EUV wave associated with a filament eruption, a CME of $\sim$520 km s$^{-1}$, a C6.6 flare, as integral components of a typical solar eruption (Fig.1). The existence of the type II radio burst confirms the formation of a coronal shock (Ginzburg \& Zhelezniakov 1958), and the estimated shock speed is $\sim$1500 km s$^{-1}$ (Fig.2). QFP wave trains originated from the core region of the flare, and propagated at a speed of $\sim$1200-1600 km s$^{-1}$ with a period of $\sim$2.5 minutes. The wave trains abruptly disappeared as they approached the CME flank (Fig.3). The EUV wave had a fast initial speed of $\sim$1000 km s$^{-1}$ (Fig.4(h)-(i)), comparable to those of the shock and QFP wave trains. The EUV wave quickly decreased to $\sim$380 km s$^{-1}$, and finally disappeared at the boundary of the CH and the north end of the streamer-like structure. The commencement of the EUV wave was near the times of the filament/CME acceleration and the flare impulsive phase, but the beginning location of the EUV wave was very close to the eruption center (Fig.4(g)-(h)). It suggests that the EUV wave was likely initiated behind the outer expanding CME front. Therefore, we believe that the EUV wave is a coronal fast-mode shock wave, and it is likely triggered by the flare impulses (Vr{\v s}nak \& Cliver 2008; Magdaleni{\'c} et al. 2012; Kumar \& Innes 2015).

Attractively, well-defined SWs ascended upwards following the incident EUV wave (Fig.5). Is it possible that the upward SWs resulted from the interaction between the EUV wave and the elongated CH? First of all, after the sweeping of the EUV wave, there were only persistent brightenings at the CH boundary (blue arrows in Fig.4(i) and Fig.5(g)), and no horizontal reflection, refraction, or transmission of the EUV wave were detected from the the boundary of the CH (Fig.4(i)). Secondly, no signal of upward waves was found from the CH interior (Fig.5(h)), and persistent brightenings indicate that no energy/wave was released at the CH boundary. Therefore, the upward SWs are not explained as a result of the interaction between the EUV wave and the CH.

The positions of S5-S6 far away from the CH and the loop shape of S9 imply that upward SWs arise in the streamer-like structure. Because of the position near the limb and the background corona in the streamer-like structure, it is hard to locate the start and end of SWs. The SWs are traced back to limb brightenings that were induced by the earlier EUV wave. The close temporal relationship between the EUV wave and upward SWs is shown in time-space plots (Fig.5(e)-(g) and (i)). It has also been confirmed that the fast-mode MHD wave could be partly trapped inside magnetic loops rooting at magnetic QSLs and formed stationary front/brightenings (Chandra et al. 2016; Chen et al. 2016, Yuan et al. 2016; Zong \& Dai 2017). On the other hand, the interaction between the shock and the streamer-like structure is shown by the sudden change of the radio spectrum (the yellow box in Fig.2(d)). The lower start position (1.18 $R_\odot$) of the shock can make trapped waves in the lower corona be released. The brightenings both on the limb and at the CH boundary possibly resulted from localized energy release triggered by the EUV wave (Ofman \& Thompson 2002; Kwon et al. 2013). The brightenings at the CH boundary were persistent, but those on the limb faded away as soon as SWs set off (Fig.5(e)-(g) and (i)). The speed ($\sim$80 km s$^{-1}$) of upward SWs is in the speed range of slow-mode waves. Hence, we propose that the upward SWs likely originated from the partially trapped fast-mode EUV wave, and were possibly activated by the strong compression of the shock wave in the form of slow-mode. %On the other hand, the spectral bump of the type II radio burst also manifests the interaction between the shock wave and streamer-like structure (Feng et al. 2012).

The configuration of the streamer-like structure is important to trap the part of the EUV wave, however upward SWs are barely detected before. The strong compression of the lower shock on the streamer-like structure should be crucial to release the trapped lower EUV wave in the form of upward slow-mode SWs.

\acknowledgments
SDO is a mission of NASA's Living With a Star Program. The authors thank the teams of SDO, SOHO, GONG, YNAO, and LEAR for providing the data. This work is supported by grants NSFC 41331068, 11303101, and 11603013, Shandong Province Natural Science Foundation ZR2016AQ16, and Young Scholars Program of Shandong University, Weihai, 2016WHWLJH07. H. Q. Song is supported by the Natural Science Foundation of Shandong Province JQ201710.

\clearpage

\begin{figure}
\epsscale{0.95} \plotone{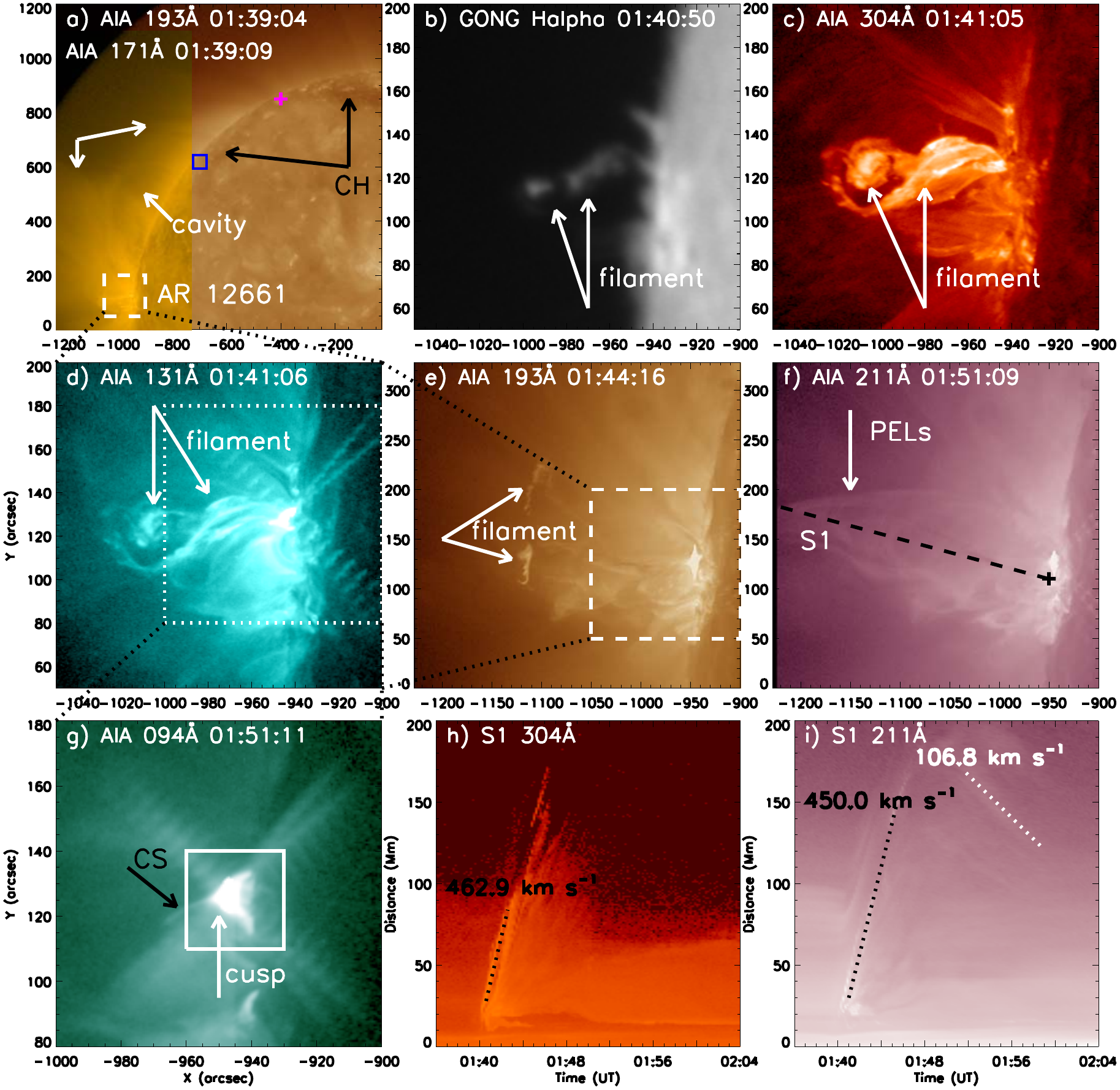}
\caption{The filament eruption in the GONG H$\alpha$ filtergram and AIA EUV channels. (a)The elongated CH (black arrows), coronal cavity and overlying streamer-like structure (white arrows) to the north of AR 12661. (b-g) The twist structure of the erupting filament, PELs, the cusp, and the CS. (h-i) Time-space plots of AIA 304 and 211~{\AA} images showing the evolution of the erupting filament and PELs along the S1 (the dashed line starting from a plus in (f)). The dotted lines are used to derive the associated speeds. The FOV of (b)-(d) is indicated by the dashed boxes in (a) and (e), and the dotted box in (d) show the FOV of (g). The box in (g) is used to get the intensity curve of AIA 94~{\AA}. (An animation of the filament eruption in the GONG H$\alpha$ filtergram and seven AIA EUV channels is available in the online Journal. The animated sequence runs from 01:30 to 2:00 UT.)
\label{f1}}
\end{figure}

\clearpage

\begin{figure}
\epsscale{1.0}
\plotone{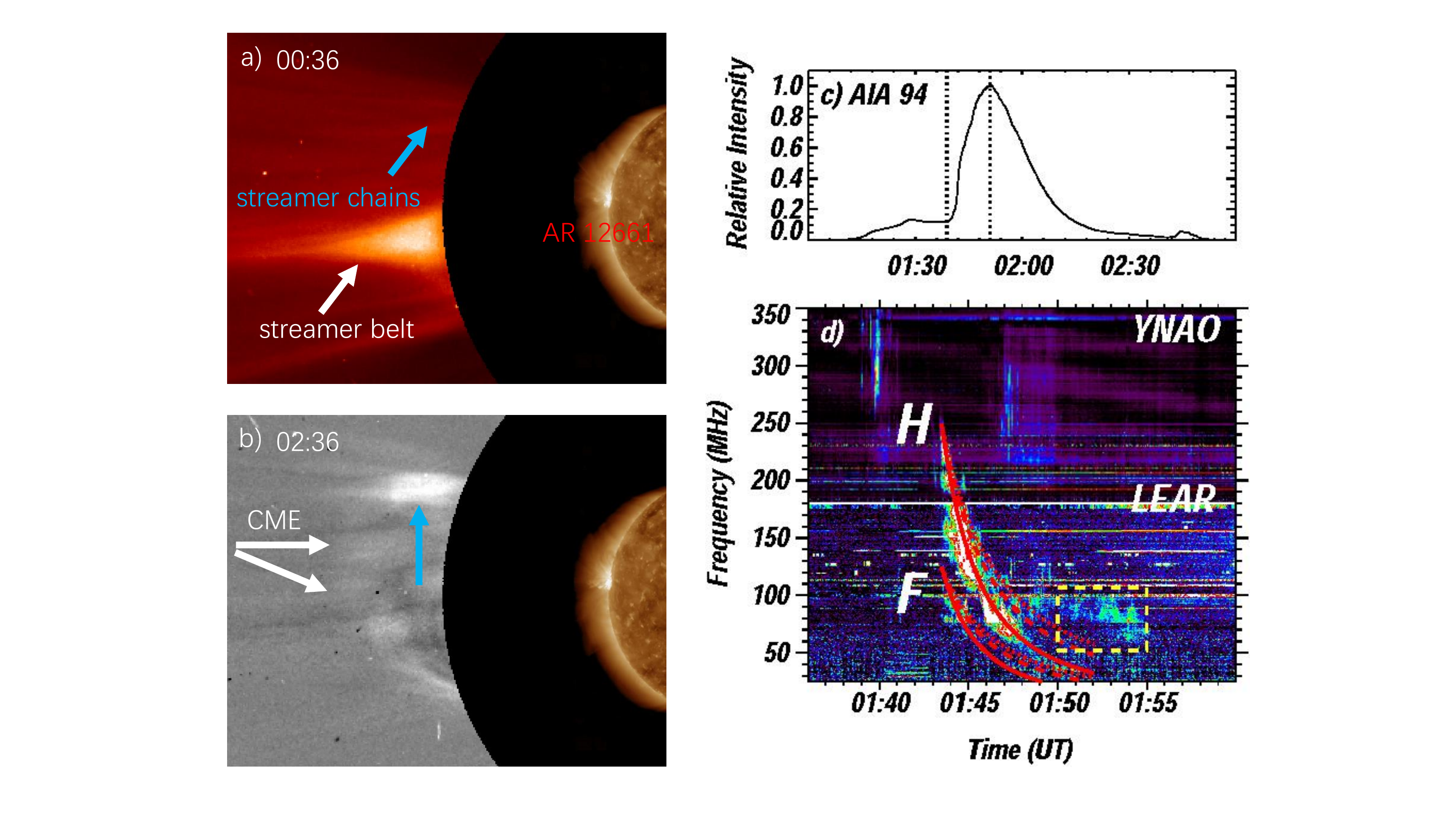}
\caption{(a-b) The CME emanating below the streamer belt (white arrows) and the disturbance in streamer chains (blue arrows) in composite images of LASCO C2 and AIA 193~{\AA}. (c) The intensity curve of the box in Fig.1(g) in AIA 94~{\AA}, and the dotted lines mark the beginning and peak times of the flare impulsive phase. (d) Composition radio dynamic spectrum obtained by LEAR (25-180 MHz) and YNAO (180-350 MHz). Red curves are fittings to the fundamental (F) and harmonic (H) branches with an assuming shock speed of 1500 km s$^{-1}$, using the one-fold (solid), three-fold (dashed), and five-fold (dotted) Newkirk models, respectively. The dashed box encircles the sudden change of the radio spectrum. (A color figure is available online.)
\label{f2}}
\end{figure}

\clearpage

\begin{figure}
\epsscale{0.95}
\plotone{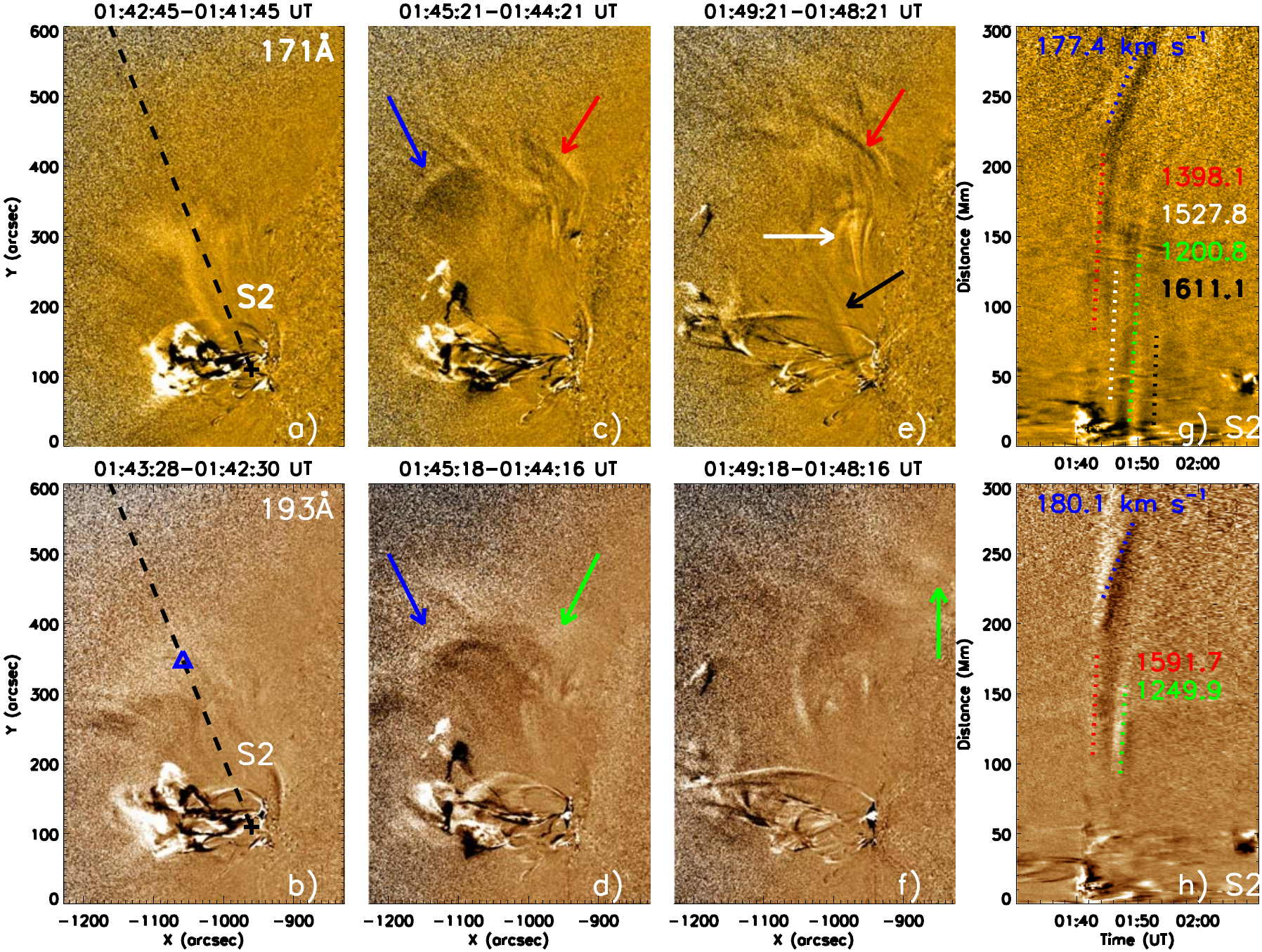}
\caption{(a-f) Running-ratio-difference images in AIA 171 and 193~{\AA} showing the wave trains along the loops (the black arrow). The blue and green arrows indicate the CME front and the EUV wave, respectively. The white and red arrows show the bright top of the coronal cavity and the displacement of the overlying loops. (g-h) Time-space plots of running-ratio-difference AIA 171 and 193~{\AA} images uncovering the CME front (blue) and the wave trains (red, white, green, and black) along the dashed lines (S2) in (a)-(b). The dotted lines are used to derive the attached speeds. (An animation of the running-ratio-difference images is available in the online Journal. The animated sequence runs from 01:30 to 2:00 UT.)
\label{f3}}
\end{figure}

\clearpage

\begin{figure}
\epsscale{0.9} \plotone{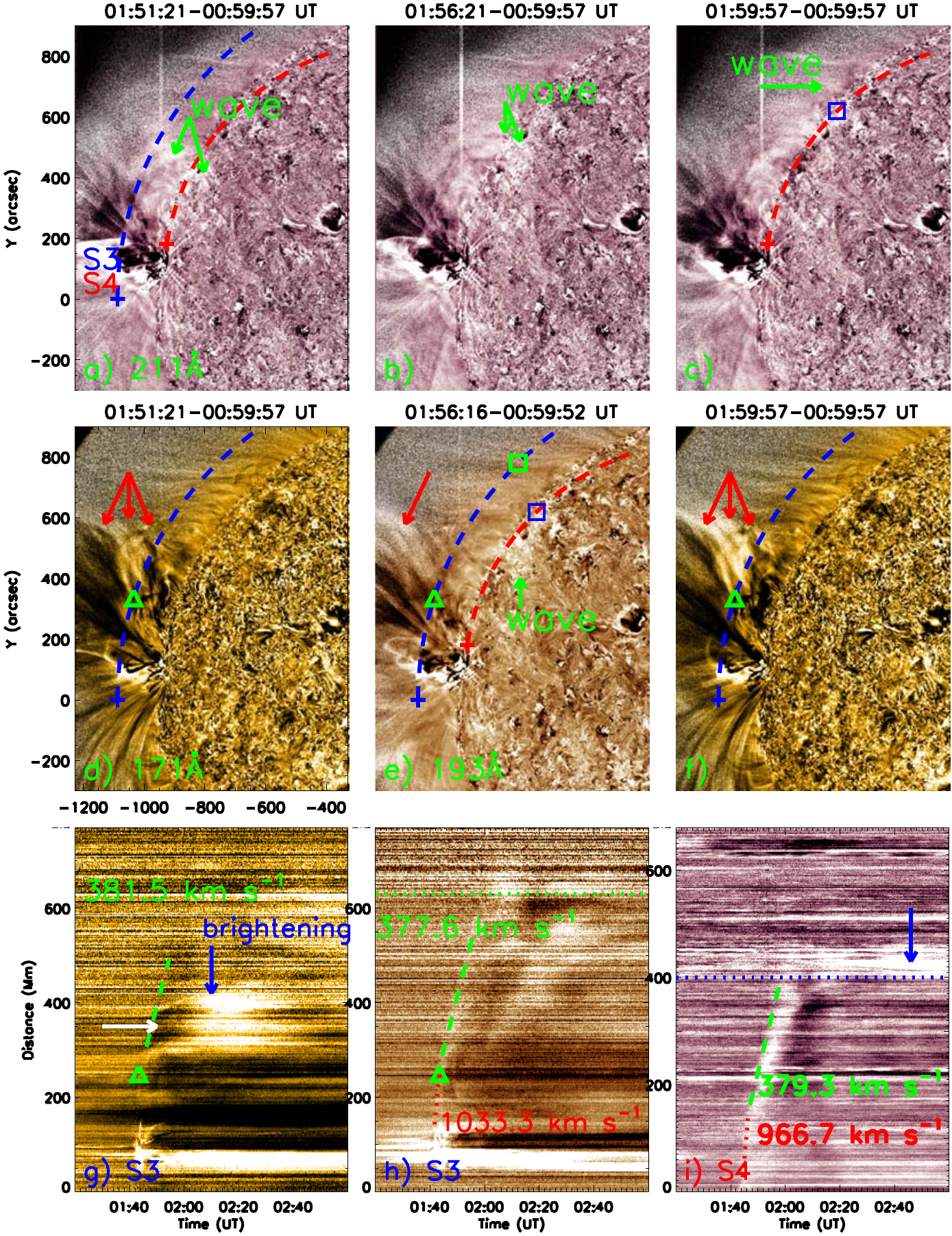}
\caption{(a-f) Base-ratio-difference images in AIA 211, 171 and 193~{\AA} showing the EUV wave (green arrows) and the associated lightened loops (red arrows). (g-i) Time-space plots of base-ratio-difference images in AIA 171, 193, and 211~{\AA} displaying the wave propagation and brightenings (the blue arrow) along the dashed arcs (S3-S4) in (a). The green triangles in (d)-(h) represent the deceleration position of the EUV wave. The horizontal lines and squares indicate the terminal points of the EUV wave along S3 (green) and S4 (blue), respectively. The inclined dashed and dotted lines are used to derive the attached speeds. (An animation of the base-ratio-difference images is available in the online Journal. The animated sequence runs from 01:40 to 2:38 UT.)
\label{f4}}
\end{figure}

\clearpage

\begin{figure}
\epsscale{1.0} \plotone{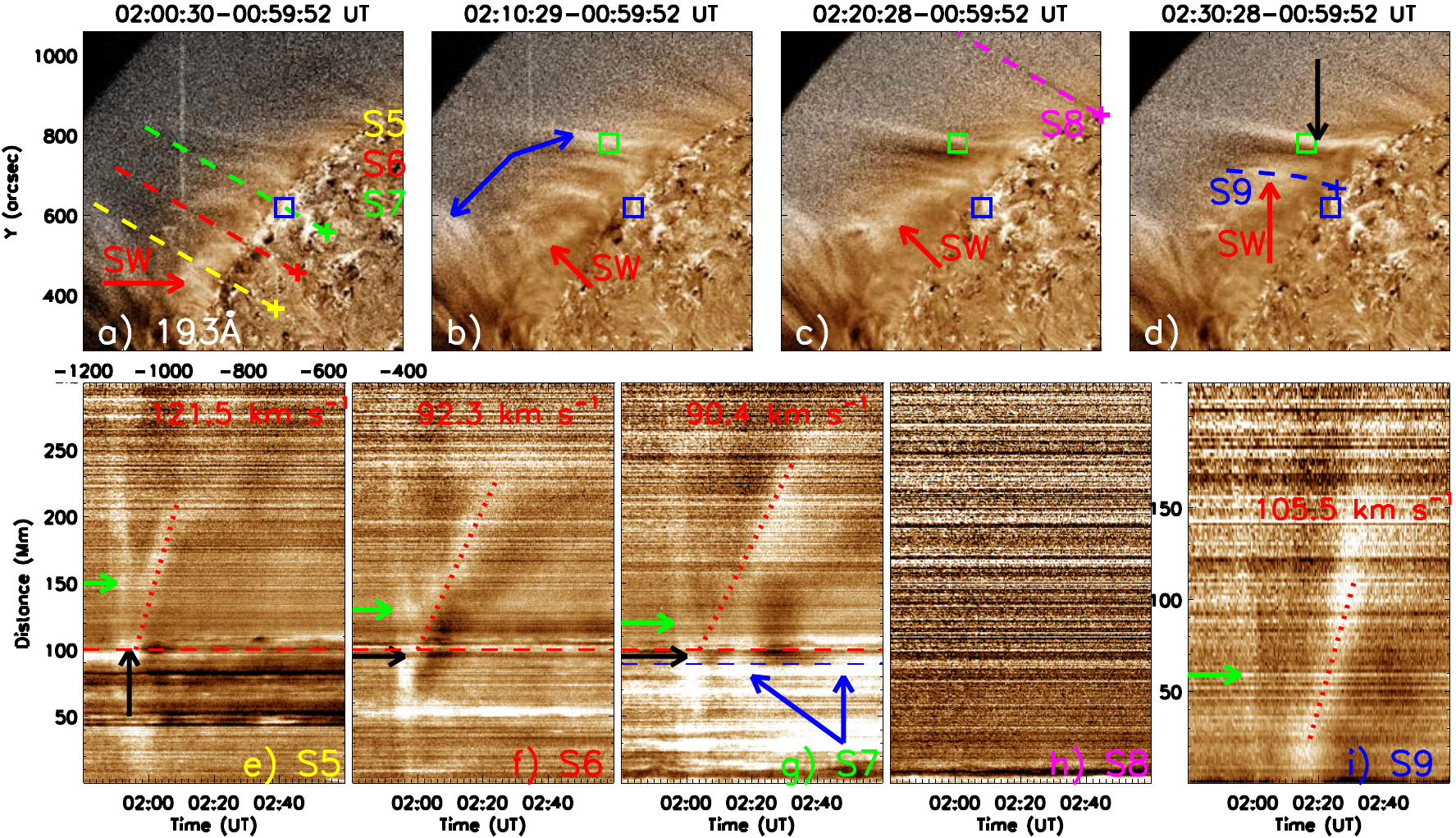}
\caption{(a-d) Base-ratio-difference AIA 193~{\AA} images showing upward SWs (red arrows) in the streamer-like structure (blue arrows). The black arrow indicates the lightened loops. (e-i) Time-space plots of base-ratio-difference AIA 193~{\AA} images showing the upward SWs, the EUV wave (green arrows), and brightenings (black and blue arrows) along the dashed lines (S5-S9) in upper panels. The dotted lines are used to derive the attached speeds, and the red and blue horizontal lines mark the solar limb and the CH boundary, respectively. (An animation of the base-ratio- difference AIA 193 ?A images is available in the online version of Figure 4.)
\label{f5}}
\end{figure}

\end{document}